\documentclass[journal=jctcce,manuscript=article,layout=twocolumn]{achemso}
\setkeys{acs}{articletitle=true}

\usepackage{chemformula} % Formula subscripts using \ch{}
\usepackage[T1]{fontenc} % Use modern font encodings
\usepackage{amsmath,amsfonts}
\usepackage{physics}
\usepackage{comment}
\usepackage{booktabs}
\usepackage{multirow}
\usepackage{textcomp}  % \textdegree symbol
\mciteErrorOnUnknownfalse
\usepackage{subfigure}

\renewcommand{\vec}[1]{\boldsymbol{#1}}
\newcommand*\spanfun[1]{\,\mathrm{span}\bigl\{#1\bigr\}}
\DeclareMathOperator*{\argmax}{arg\,max}
\DeclareMathOperator*{\argmin}{arg\,min}

\newcommand{\ceil}[1]{\lceil #1 \rceil}

\newcommand*\bigO{O}

\newcommand*\fname{coordinate descent FCI}
\newcommand*\Fname{Coordinate descent FCI}
\newcommand*\sname{CDFCI}
\newcommand*\fciqmc{iS-FCIQMC}

\newcommand*\bbR{\mathbb{R}}
\newcommand*\calI{\mathcal{I}}
\newcommand*\calV{\mathcal{V}}

\newcommand*\itell{{(\ell)}}
\newcommand*\itellp{{(\ell+1)}}

\newcommand*\varalpha{\widetilde{\alpha}}
\newcommand*\nelec{n_{\text{e}}}
\newcommand*\norb{n_{\text{orb}}}
\newcommand*\hset[1]{\calI_H(#1)}

\newcommand*\wds[1]{{\it #1}}

\author{Zhe Wang}
\author{Yingzhou Li}
\email{yingzhou.li@duke.edu}
\author{Jianfeng Lu}
\email{jianfeng@math.duke.edu}
\affiliation[Duke]{Department of Mathematics, Duke University}
\alsoaffiliation{Department of Chemistry and Department of Physics,
Duke University}

\title[Coordinate Descent FCI]
{Coordinate Descent Full Configuration Interaction}

\abbreviations{FCI, CDM, \sname{}}

\begin{document}

% \begin{tocentry}
% 
%     Some journals require a graphical entry for the Table of Contents.
%     This should be laid out ``print ready'' so that the sizing of the
%     text is correct.
% 
%     Inside the \texttt{tocentry} environment, the font used is Helvetica
%     8\,pt, as required by \emph{Journal of the American Chemical
%     Society}.
% 
%     The surrounding frame is 9\,cm by 3.5\,cm, which is the maximum
%     permitted for  \emph{Journal of the American Chemical Society}
%     graphical table of content entries. The box will not resize if the
%     content is too big: instead it will overflow the edge of the box.
% 
%     This box and the associated title will always be printed on a
%     separate page at the end of the document.
% 
% \end{tocentry}

\begin{abstract}

  We develop an efficient algorithm, \fname{} (\sname{}), for the
  electronic structure ground state calculation in the configuration
  interaction framework. \sname{} solves an unconstrained non-convex
  optimization problem, which is a reformulation of the full
  configuration interaction eigenvalue problem, via an adaptive
  coordinate descent method with a deterministic compression
  strategy. \sname{} captures and updates appreciative determinants
  with different frequencies proportional to their importance. We show
  that \sname{} produces accurate variational energy for both static
  and dynamic correlation by benchmarking the binding curve of
  nitrogen dimer in the cc-pVDZ basis with $10^{-3}$ mHa accuracy.  We
  also demonstrate the efficiency and accuracy of \sname{} for
  strongly correlated chromium dimer in the Ahlrichs VDZ basis and
  produces state-of-the-art variational energy.

\end{abstract}

\section{Introduction}

Solving quantum many-body problem for electrons is a well-known
challenging task. While weakly correlated (single-reference) systems
can be well approximated using density functional theory and coupled
cluster methods such as CCSD(T); strongly corrected (multi-reference)
systems remain challenging.  The difficulty comes in two aspects: the
infamous fermion sign problem and combinatorial scaling of the problem
size. In this paper, we propose an efficient algorithm, named \fname{}
(\sname{}), to calculate the ground state energy and its corresponding
variational wavefunction for both weakly and strongly correlated
fermion systems in the framework of full configuration interaction.

Besides the direct diagonalization of the full configuration
interaction (FCI) Hamiltonian~\cite{Knowles1984}, many other
algorithms have been proposed, which can be roughly organized into
three groups. The first group, density matrix renormalization group
(DMRG)~\cite{White1999, Chan2011, Olivares-Amaya2015}, adopts tensor
train ansatz in representing the variational wavefunction. DMRG has
been routinely applied to study the ground and excited state of
strongly correlated $\pi$-conjugated molecules and one-dimensional
systems~\cite{Olivares-Amaya2015}.  The second group, like full
configuration interaction quantum Monte Carlo
(FCIQMC)~\cite{Booth2009, Booth2012, Lu2017}, assumes that the ground
state variational wavefunction can be represented as the empirical
distribution of a large number of stochastic walkers. To reduce the
variance of the energy estimator and the required number of walkers,
initiator-FCIQMC (iFCIQMC)~\cite{Cleland2010} and semi-stochastic
FCIQMC (S-FCIQMC)~\cite{Petruzielo2012} are developed aiming at a good
trade-off between variance and bias.  The third group first solves a
selected configuration interaction (SCI) problem and then conducts a
perturbation calculation.  This family of algorithms (SCI+PT),
includes the early work on configuration interaction by perturbatively
selecting iteration (CIPSI)~\cite{Huron1973}, and more recently,
adaptive configuration interaction (ACI)~\cite{Schriber2017}, adaptive
sampling configuration interaction (ASCI)~\cite{Tubman2016},
\latin{etc}.  Heat-bath configuration interaction
(HCI)~\cite{Holmes2016} significantly reduces the computational cost
of the selected CI phase based on the information from magnitudes of
the double excitations. With perturbation, HCI is able to calculate
the ground state energy of a strongly correlated all electron chromium
dimer up to $1$~mHa accuracy in Ahlrichs VDZ basis. More recently,
semi-stochastic HCI (SHCI)~\cite{Sharma2017} further accelerates the
perturbation phase with a stochastic idea similar to FCIQMC.

\begin{figure}[ht]
    \centering
    \includegraphics[width=0.5\textwidth]{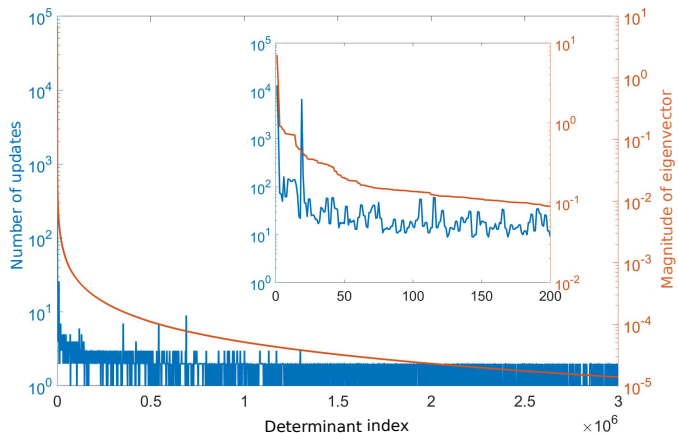}
    \caption{Correlation between the updating frequencies and magnitudes
      of the ground state wavefunction of an all-electron \ch{C2} with
      cc-pVDZ basis calculated by \sname. Coefficients of configuration
      interaction wavefunction are sorted in a decreasing order based on
      their magnitudes. Smaller panel shows the results of the largest
      200 coefficients.}
    \label{fig:updatefreq}
\end{figure}

The algorithm we considered in this paper belongs to the third group
(SCI+PT) above. Our goal is to improve the variational stage of the
computation \latin{i.e.}, the selective CI part. Our proposed
algorithm, \sname{}, is an adaptive coordinate-wise (\latin{i.e.},
determinant-wise) iterative method. It updates the coefficient of an
appreciative determinant each iteration and has the nice feature of
visiting determinants with different frequencies proportional to their
importance. As not all determinants contribute equally to the ground
state wavefunction, \sname{} is able to efficiently capture the
important part of the FCI space and obtain a good approximation to the
ground state. Figure~\ref{fig:updatefreq} indicates the relation
between the updating frequency of determinants and magnitudes of the
determinant coefficients of the ground state wavefunction for an
all-electron \ch{C2} calculation with cc-pVDZ basis by \sname{}. As
shown in Figure~\ref{fig:updatefreq}, many coefficients are only
updated once throughout iterations, which shows the efficiency of the
updating strategy in \sname{}. During iterations, \sname{} also
compresses those unappreciative determinants through hard
thresholding. Our implementation philosophy of \sname{} is to reserve
memory resource for storing variational wavefunction as much as
possible, hence, the Hamiltonian matrix is evaluated on-the-fly.
Eventually, \sname{} is able to capture the binding curve of
all-electron \ch{N2} with cc-pVDZ basis up to 6 digits accuracy in one
week and compute the ground state energy of all-electron \ch{Cr2} with
Ahlrichs VDZ basis to $-2086.443565$ Ha, which is the state-of-the-art
variational result.

The rest of the paper is organized as follows. Section~\ref{sec:algo}
presents the \sname{} algorithm. The implementation detail is stated in
Section~\ref{sec:impl}. In Section~\ref{sec:numres}, we demonstrate the
efficiency and accuracy of \sname{} via applying it to various molecules
including \ch{H2O}, \ch{C2}, \ch{N2}, and \ch{Cr2}. Also the binding
curve of \ch{N2} is characterized. Finally, in Section~\ref{sec:conc},
we conclude the paper together with discussion on future work.

\section{\Fname{}}
\label{sec:algo}

This section first reformulates the FCI eigenvalue problem as a
non-convex optimization problem~\cite{Lei2016, Li2018b} with no spurious
local minima and then describes in detail the \fname{} algorithm together
with the compression technique and suggested stopping criteria.

Given a complete set of spin-orbitals $\{\chi_p\}$, a many-body
Hamiltonian operator, under the second quantization, can be written as
\begin{equation} \label{eq:second_quantized_h}
    \widehat{H} = \sum_{p,q} t_{pq} \hat{a}_{p}^\dagger \hat{a}_{q}
    + \frac{1}{2} \sum_{p,r,q,s} v_{prqs} \hat{a}_{p}^\dagger
    \hat{a}_{r}^\dagger \hat{a}_{q} \hat{a}_{s},
\end{equation}
where $\hat{a}_{p}^\dagger$ and $\hat{a}_{p}$ denote the creation
and annihilation operator of an electron with spin-orbital index
$p$, $t_{pq}$ and $v_{prqs}$ are one- and two-electron integrals
respectively. The ground state energy of $\widehat{H}$ can be obtained
from solving the time-independent Schr\"{o}dinger equation
\begin{equation} \label{eq:eig}
  \widehat{H} \ket{\Phi_0} = E_0 \ket{\Phi_0},
\end{equation}
where $E_0$ denotes the ground state energy (the smallest eigenvalue)
and $\ket{\Phi_0}$ denotes the corresponding ground state
wavefunction.  Without loss of generality, we assume that $E_0$ is
negative and non-degenerate, i.e., the eigenvalues of $\widehat{H}$
are given as $E_0 < E_1 \leq E_2 \leq \cdots$.

In FCI, the complete spin-orbital set is truncated to a finite subset
$\{\chi_p\}_{p = 1}^{\norb}$, obtained e.g., by Hartree-Fock
or Kohn-Sham calculations. The FCI variational space $\calV$ is
spanned by all possible Slater determinants constructed from
$\{\chi_p\}_{p = 1}^{\norb}$, and the dimension of $\calV$ is
denoted as $N_{\text{FCI}}$ also known as the total number of
configuration interactions.  The ground state wavefunction is then
discretized in $\calV$, \latin{i.e.},
$\ket{\Phi_0} \in \calV = \spanfun{\ket{D_1}, \dots,
  \ket{D_{N_\text{FCI}}}}$, where $\ket{D_1}$ denotes the reference
determinant, $\ket{D_i}$ for $1 < i \leq N_\text{FCI}$ denotes other
Slater determinants constructed from the finite set of spin-orbitals.
Correspondingly, the Hamiltonian operator is represented by a
many-body Hamiltonian matrix $H$ with its $(i,j)$ entry as
$H_{i,j} = \bra{D_i}\widehat{H}\ket{D_j}$. Let $\vec{b}$ and $\vec{c}$
denote coefficient vectors with entry $b_i$ and $c_i$ respectively.
The ground state wavefunction can be written as
$\ket{\Phi_0} = \sum_i c_i \ket{D_i}$.  The time-independent
Schr\"odinger equation \eqref{eq:eig} has its matrix representation
as,
\begin{equation}
  \label{eq:eig-fci}
   H \vec{c} = E_0 \vec{c},
\end{equation}
which is known as the FCI eigenvalue problem.  The second-quantized
Hamiltonian operator as in \eqref{eq:second_quantized_h} implies that
$H_{i,j}$ is nonzero if and only if $\ket{D_i}$ can be obtained from
$\ket{D_j}$ via changing at most two occupied spin-orbitals. Hence, we
say that $\ket{D_i}$ is $H$-connected with $\ket{D_j}$ if $H_{i,j}$
is nonzero. The set of all indices $i$ such that $\ket{D_i}$ is
$H$-connected with $\ket{D_j}$ is called the $H$-connected index set of
$j$ and is denoted as $\hset{j}$.  Since the cardinality of $\hset{j}$ is
much smaller than $N_{\text{FCI}}$, the matrix $H$ is extremely sparse.

\subsection{Reformulation of the FCI eigenvalue problem}

The FCI eigenvalue problem \eqref{eq:eig-fci} can be reformulated as the
following unconstrained non-convex optimization problem,
\begin{equation} \label{eq:opt}
    \min_{\vec{c} \in \bbR^{N_{\text{FCI}}}} f(\vec{c}) = \norm{H +
    \vec{c}\vec{c}^\top}_F^2,
\end{equation}
where $\norm{\cdot}_{F}$ denotes the Frobenius norm of a matrix.
The gradient of the objective function is
\begin{equation*}
    \nabla f(\vec{c}) = 4H \vec{c} + 4 \left( \vec{c}^\top \vec{c}
    \right) \vec{c}.
\end{equation*}
As analyzed in our previous work~\cite{Li2018b}, the stationary points
are $\vec{0}$ and $\pm \sqrt{-E_i} \vec{v}_i$ for $E_i < 0$, where
$\vec{v}_i$ is the normalized eigenvector corresponding to $E_i$.
Furthermore, importantly, $\pm \sqrt{-E_0} \vec{v}_0$ are the only two
local minimizers with the same objective value, while other stationary
points are saddle points. Thus, solving the optimization problem
\eqref{eq:opt} reveals the ground state energy $E_0$ and the
ground state wavefunction coefficient vector $\vec{v}_0$. For such an
optimization problem, higher order methods converge to global minima
efficiently but their per iteration computational costs are too
expensive for FCI problems.  Hence, only first order methods, like
gradient descent methods (GDs), stochastic gradient descent methods
(SGDs), and coordinate descent methods (CDMs), are discussed
here.

The first order optimization methods applied to \eqref{eq:opt},
compared with solving \eqref{eq:eig-fci} using traditional methods,
\latin{e.g.}, power method, Davidson method, and Lanczos method, have
two main advantages. First, GDs, SGDs, and CDMs do not need any tuning
parameters: no diagonal shift is needed to turn the smallest
eigenvalue into the largest one in magnitude~\cite{Li2018b} and the
stepsize can be addressed by an exact line search. Second, since no
orthonormality constraint appears explicitly in \eqref{eq:opt},
solving \eqref{eq:opt} with GDs, SGDs, and CDMs does not need any
orthonormalization step. While in traditional methods like Davidson or
Lanczos methods, orthonormalization step is needed every a few
iterations to avoid numerical instability issue, which would be
expensive for FCI problems.

Among the first order optimization methods, CDMs are more suitable
for FCI problems. GDs evaluate and update the exact gradient each
iteration, which is prohibitively expensive for FCI problems,
due to the huge problem dimension. SGDs evaluate and update a
stochastic approximation of the full gradient, which is of much
cheaper computational cost per iteration. SGDs actually have been
applied to FCI problems in an implicit way: as FCIQMC, iFCIQMC, and
S-FCIQMC can all be regarded as SGDs applied to a similar objective
function as \eqref{eq:opt} with constant stepsizes~\cite{Lu2017}. SGDs,
in general, converge efficiently to a neighborhood of minimizers and then
wander around the minimizer due to the stochastic approximation. CDMs
select and update a single coefficient each iteration, which is of
cheap computational cost.  Comparing to GDs, CDMs provably achieve
faster convergence rate in terms of the prefactor~\cite{Li2018b},
whereas comparing to SGDs, CDMs are approximately of equal cost
per iteration, but is much more stable towards convergence. Further,
with a properly designed selecting strategy, CDMs updates different
coefficients with different frequencies, taking advantage of the
different importance of determinants in FCI problems.  Taking these
advantages into consideration, we design \sname{}, which is a CDM
tailored for FCI problems with compression strategy, to efficiently
solve \eqref{eq:opt}. This method is described in details below.

\subsection{Algorithm}

The \sname{} algorithm stores two sparse vectors $\vec{c}^\itell$ and
$\vec{b}^\itell$ in the main memory that are contiguously maintained
throughout iterations: $\vec{c}^\itell$ denotes the computed coefficient
vector of the ground state wavefunction in the $\ell$-th iteration and
$\vec{b}^\itell$ is a compressed approximation of $H\vec{c}^\itell$.

Let us first give a sketch of the \sname{} algorithm: In the $\ell$-th
iteration, \sname{} first finds the $i^\itellp$-th determinant with
potentially steepest objective function value decrease among all
$H$-connected determinants of $\ket{D_{i^\itell}}$.  Then \sname{}
conducts an exact line search to find the optimal update $\alpha$ such
that $f(\vec{c}^\itell + \alpha \vec{e}_{i^\itellp})$ is minimized
and hence the estimator for the ground state energy is reduced, where
$\vec{e}_{i^\itellp}$ denotes the indicator vector of $i^\itellp$.
$\vec{b}^\itell$ plays an important role in all above steps. In
preparation for next iteration, the vector $\vec{b}^\itellp$ needs
to be updated incorporating with $\vec{c}^\itellp = \vec{c}^\itell +
\alpha \vec{e}_{i^\itellp}$. However, an exact update $\vec{b}^\itellp
= \vec{b}^\itell + \alpha H_{:,i^\itellp}$ could waste limited memory
resource on unappreciative determinants.  Our compression step updates
coefficients only when they do not cost extra memory resource or if they
are significantly large in magnitudes. Although the compression step
introduces error along the iterations, as we will show, we can still
calculate the Rayleigh quotient corresponding to $\vec{c}^\itellp$
exactly, which is used as the ground state energy estimator in \sname{}.

In the following, we will discuss each part of \sname{} in detail and
then conclude this section with a pseudo-code for the algorithm.

\subsubsection{Determinant-select and coefficient-update}
\label{sec:coord-select}

Determinant-select is the first step in each iteration. Assume that the
$\ell$-th iteration updates determinant $\ket{D_{i^\itell}}$ and results
in a coefficient vector $\vec{c}^\itell$. We select the determinant to
be updated at the current iteration, $\ket{D_{i^\itellp}}$, according
to local information at $\vec{c}^\itell$. In order to decrease the
objective function value to the largest extent, we could select the
determinant with the largest magnitude of the approximated gradient at
$\vec{c}^\itell$, \latin{i.e.}, $\ket{D_{i^\itellp}}$ with
\begin{equation*}
    i^\itellp = \argmax_{j} \abs{4 b^\itell_j + 4 \left(
    \left(\vec{c}^\itell\right)^\top \vec{c}^\itell \right)
    c_j^\itell},
\end{equation*}
where $\vec{b}^\itell$ is a compressed approximation of
$H \vec{c}^\itell$.  However, the above strategy requires checking
each $j$ (i.e., all determinants), which is prohibitive for even
moderate size problems. Hence, instead of checking all determinants,
\sname{} only checks the $H$-connected determinants of
$\ket{D_{i^\itell}}$, \latin{i.e.},
\begin{equation} \label{eq:coord-select}
    \begin{split}
        & i^\itellp = \\
        & \argmax_{j \in \hset{i^\itell}}
        \abs{4b_j^\itell +
        4 \left( \left( \vec{c}^\itell \right)^\top \vec{c}^\itell \right)
        c_j^\itell}.
    \end{split}
\end{equation}
Since our compression strategy introduced later in Section~\ref{sec:comp}
truncates unappreciative determinants, the expression in
\eqref{eq:coord-select} remains a good approximation of the exact
gradient at $\vec{c}^\itell$. Empirically, such a gradient-based
determinant-select strategy outperforms other perturbation-based
determinant-select strategies as used in other SCI algorithms (see
Section~\ref{sec:numres} for details).

Once the $i^\itellp$-th determinant is selected, \sname{} determines
the stepsize by the line search along that direction so to decrease the
objective function value by the largest amount. Denoting the update as
$\alpha$, the line search can be formulated as
\begin{equation} \label{eq:coord-update-opt}
    \alpha = \argmin_{\varalpha \in \bbR} f(\vec{c}^\itell +
    \varalpha \vec{e}_{i^\itellp}).
\end{equation}
Since $h(\varalpha) = f(\vec{c}^\itell + \varalpha \vec{e}_{i^\itellp})$
is a quartic polynomial in $\varalpha$, solving the minimization
problem \eqref{eq:coord-update-opt} is equivalent to finding roots of
$h'(\varalpha)$ -- the derivative of $h(\varalpha)$. If $h'(\varalpha)$
has a unique root, then the root is the minimizer. If $h'(\varalpha)$
has two roots, then the one with multiplicity one is the minimizer. If
$h'(\varalpha)$ has three roots, then the one further away from the
middle one is the minimizer. Given the update $\alpha$, we can easily
update $\vec{c}^\itell$ as
\begin{equation} \label{eq:coord-update-c}
    c_{i}^\itellp =
    \begin{cases}
        c_i^\itell, & i \neq i^\itellp;\\
        c_i^\itell + \alpha, & i = i^\itellp.
    \end{cases}
\end{equation}

In \sname{}, we also need to maintain $\vec{b}^\itellp \approx H
\vec{c}^\itellp$ for future determinant-select steps. Since only
one coefficient is updated in $\vec{c}^\itell$, the corresponding
$\vec{b}^\itell$ can be updated accordingly as,
\begin{equation} \label{eq:coord-update-b}
    \vec{b}^\itellp \approx H \vec{c}^\itellp \approx \vec{b}^\itell +
    \alpha H_{:,i^\itellp}.
\end{equation}
Therefore, each update step requires evaluation of all $H$-connections
from $\ket{D_{i^\itellp}}$. Besides the update from $\vec{c}^\itellp$, we
also recalculate the current $i^\itellp$-th entry in $\vec{b}^\itellp$
to guarantee the correctness and increase the numerical stability
of our algorithm. Such a recalculation could improve the accuracy
of the determinant-select \eqref{eq:coord-select} and line search
\eqref{eq:coord-update-opt} in the following iterations, and also
provide an accurate Rayleigh quotient as the estimator of the ground
state energy as in \eqref{eq:coord-update-cHc}.  We argue this correction
comes for free in addition to \eqref{eq:coord-update-b}, since
\begin{equation} \label{eq:coord-update-bj}
    \begin{split}
        b_{i^\itellp}^\itellp = & H_{i^\itellp,:} \vec{c}^\itellp \\
    = & \sum_{j \in \hset{i^\itellp}} (H_{j,i^\itellp})^\ast
        c_j^\itellp,
    \end{split}
\end{equation}
where $(H_{j,i^\itellp})^\ast$ denotes the complex conjugate of
$H_{j,i^\itellp}$, which has already been evaluated when updating
\eqref{eq:coord-update-b}.

\subsubsection{Coefficient compression}
\label{sec:comp}

Since \sname{} initializes $\vec{c}^{(0)}$ with the reference determinant
$\ket{D_1}$ and $\vec{b}^{(0)} = H \vec{c}^{(0)}$, the coefficient
of the reference determinant in $\vec{b}^{(0)}$ is nonzero and the
reference determinant is in $\vec{b}^{(0)}$. In later iterations,
\sname{} is designed to follow one rule: if a determinant is in
$\vec{c}^\itell$, then it is in $\vec{b}^\itell$ as well.  Under this
rule, if a determinant $\ket{D_j}$ is neither in $\vec{c}^\itell$
nor in $\vec{b}^\itell$, according to \eqref{eq:coord-select}, this
determinant has zero value therein and will not be selected. Hence
\eqref{eq:coord-select} selects either a new determinant not in
$\vec{c}^\itell$ but in $\vec{b}^\itell$ or an old determinant already
in both $\vec{c}^\itell$ and $\vec{b}^\itell$. In \sname{}, thus, the
compression strategy compresses only the unappreciative determinants
in $\vec{b}^\itell$ to control the computation and memory cost,
which in turn restricts the growth of the coefficient vector
$\vec{c}^\itell$.

Detailed compression strategy is as follows. When a coefficient
$\alpha H_{j,i^\itellp}$ is added to $b^\itell_j$, we use a
predefined tolerance $\varepsilon$ to compress the update. If the $j$-th
determinant is already selected before, then $\alpha H_{j,i^\itellp}$
is added to $b^\itell_j$ without any compression.  If the $j$-th
determinant has not been selected in $\vec{b}^\itell$, but the update
is quantitatively large, \latin{i.e.}, $\abs{\alpha H_{j,i^\itellp}} >
\varepsilon$, it indicates that the $j$-th determinant is appreciable
and the $j$-th determinant will be added to $\vec{b}^\itell$ with
the coefficient $\alpha H_{j,i^\itellp}$. Otherwise, the update is
truncated, i.e., the coefficient in $\vec{b}^{\itell}$ remains $0$.
The described compression strategy is deterministic and satisfies the
rule that determinants with nonzero coefficients in $\vec{c}^\itell$
are in $\vec{b}^\itell$ as well. For molecules, such a deterministic
strategy outperforms other strategies including stochastic compression
schemes~\cite{Lim2017} due to its effectiveness and cheap cost.

\subsubsection{Energy estimation}

Although the vector $\vec{b}^\itellp$ is compressed, we emphasize
that the Rayleigh quotient $r(\vec{c}) = \frac{\vec{c}^\top H
\vec{c}}{\vec{c}^\top \vec{c}}$ can be maintained accurately for
$\vec{c}^\itellp$, which is used in \sname{} as the estimator of the
ground state energy. First, the squared norm of $\vec{c}^\itell$ can
be updated up to numerical error, \latin{i.e.},
\begin{equation} \label{eq:coord-update-cc}
    \left( \vec{c}^\itellp \right)^\top \vec{c}^\itellp = \left(
    \vec{c}^\itell \right)^\top \vec{c}^\itell
    + 2\alpha c_{i^\itellp}^\itell + \alpha^2.
\end{equation}
An exact update can be computed for the numerator of
the Rayleigh quotient as well, \latin{i.e.},
\begin{equation} \label{eq:coord-update-cHc}
    \begin{split}
        \left( \vec{c}^\itellp \right)^\top H \vec{c}^\itellp = & \left(
        \vec{c}^\itell \right)^\top H
        \vec{c}^\itell \\
        & + 2\alpha b_{i^\itellp}^\itellp - \alpha^2
        H_{i^\itellp,i^\itellp},
    \end{split}
\end{equation}
where $b_{i^\itellp}^\itellp$ is recalculated accurately
as discussed in Section~\ref{sec:coord-select} around
\eqref{eq:coord-update-bj}. Hence this update is accurate. The Rayleigh
quotient of the updated variational wavefunction, $r(\vec{c}^\itellp)$
is the ratio of two accurately cumulated quantity and hence
accurate. Both in theoretical and numerical results, we observed that
the Rayleigh quotient is much more accurate than the projected energy
estimator~\cite{Booth2009, Booth2012, Cleland2010, Petruzielo2012},
which is $\frac{b_1^\itellp}{c_1^\itellp}$ in our notation.

Stopping criteria can be tricky for all iterative methods, including
DMRG, FCIQMC, HCI, SHCI, \latin{etc}, and is also the case for
\sname{}. Here we propose three suggestions. As for many iterative
methods, we can stop the iteration if the updated value is small. For
\sname{}, it is suggested to monitor the cumulated updated values
across a few iterations as the stopping criteria. Another stopping
criteria is based on the change of the Rayleigh quotient. Usually,
we observe monotone decay of the Rayleigh quotient before iteration
converges. Therefore, we can stop the algorithm if the decay of
the Rayleigh quotient after a few iterations is small. The third
suggestion is based on the ratio of the number of nonzero coefficients
in $\vec{b}$ and $\vec{c}$. When the algorithm converges, this ratio
converges to 1. When the ratio is close to $1$, the error introduced by
the compression slows down the convergence significantly. Hence more
iterations do not make much accuracy improvement. Mixed use of these
stopping criteria is suggested in practice.

\medskip

We conclude this section with a pseudo-code for \sname{}: 
\begin{enumerate}
    \item Initialize $\vec{c}^{(0)}$ by the reference state
    $\ket{D_1}$ with coefficient being $1$, initialize
    $\vec{b}^{(0)} = H \vec{c}^{(0)}$, and initialize $\ell = 0$.

    \item Select a determinant with the largest gradient magnitude
    according to \eqref{eq:coord-select}. Denote the selected
    determinant as $\ket{D_{i^\itellp}}$.

    \item Solve a cubic polynomial equation to obtain the optimal update
    $\alpha$ for the selected determinant. Update the $i^\itellp$-th
    coefficient as \eqref{eq:coord-update-c}.

    \item Update $b_j^\itellp = b_j^\itell + \alpha
    H_{j,i^\itellp}$ if the $j$-th determinant is already selected
    in $\vec{b}^\itell$. Otherwise, add new determinant $\ket{D_j}$ to
    $\vec{b}^\itellp$ with coefficient $\alpha H_{j,i^\itellp}$ if
    $\abs{\alpha H_{j,i^\itellp}} > \varepsilon$.  Exactly reevaluate
    $b_{i^\itellp}^\itellp$ as \eqref{eq:coord-update-bj}.

    \item Update $\left(\vec{c}^\itellp\right)^\top \vec{c}^\itellp$ and
    $\left( \vec{c}^\itellp \right)^\top H \vec{c}^\itellp$ as
    \eqref{eq:coord-update-cc} and \eqref{eq:coord-update-cHc}
    respectively. Calculate the exact Rayleigh quotient for
    $\vec{c}^\itellp$.

    \item Repeat 2-5 with $\ell \leftarrow \ell+1$ until some stopping
    criteria is achieved.
\end{enumerate}

\section{Implementation and complexity}
\label{sec:impl}

We now give some implementation details of the algorithm, focusing on
the computationally expensive parts and the numerical stability
issues.  In the end of this section, a per iteration complexity
analysis is conducted.

The indices of Slater determinants are encoded in the way that coincides
with that in the second quantization. Suppose there are $\norb$
spin-orbitals in the FCI discretization, and $\nelec$ electrons in the
system.  Then a Slater determinant is encoded as an $\norb$-bit binary
string, with each bit representing a spin-orbital. The spin-orbital
is occupied if the corresponding bit is $1$ and unoccupied if the
bit is $0$.  The $\norb$-bit binary string is stored as an array of
$64$-bit integers. Thus, $\ceil{\frac{\norb}{64}}$ integers are needed
to represent the index of a determinant.

We now focus on the implementation detail of the determinant-update
step, as it dominates the runtime.  Since the vectors $\vec{b}$ and
$\vec{c}$ are sparse and compressed in the algorithm, their entries
cannot be contiguously stored in memory. For \sname{}, we have tried
two different data structure implementations for the combined vector
(since the indices of nonzero coefficients of $\vec{c}$ are contained
in $\vec{b}$, these two vector are stored together in a single data
structure): red-black tree and hash table~\cite{Cormen2009}.

Red-black tree is a memory compact representation of the vector: Given
that $\vec{b}$ at current iteration has $n$ nonzero coefficients,
red-black tree requires $\bigO(n)$ memory. Inserting, updating and
deleting a nonzero coefficient to this red-black tree cost $\bigO(\log
n)$ operations. The drawback is that each nonzero coefficient is a
node on the tree and hence requires extra memory to store pointers,
which turns out to be more expensive comparing to the hash table.

In \sname{}, therefore, we prefer to use a fixed-size open addressing
hash table.  The hash function mapping a configuration string to an
array index is chosen as
\begin{equation}
    \mathrm{Hash}(\vec{d}) = \vec{s} \cdot \vec{d} \; ( \mathrm{mod}\;p),
\end{equation}
where the size of the hash table is chosen to be a large prime number
$p$; $\vec{d}$ is the vector of $\ceil{\frac{\norb}{64}}$ integers with
bits representing the configuration of the determinant; and $\vec{s}$
is a fixed vector of the same length as $\vec{d}$ with entries randomly
chosen from $[0, p-1]$ during the \sname{} initialization step.  In our
current implementation, for each execution of the algorithm, we allocate
an array of size approaching machine memory limit for the hash table,
which could be modified to enable dynamic resizing feature in order to be
memory compact.  Inserting, updating and deleting a nonzero coefficient
in hash table cost $\bigO(1)$ operations on average; while in the worst
case, when the table is almost full, inserting and deleting operation
would cost $\bigO(p)$ operations.  In order to avoid such inefficient
scenarios, we limit the load factor below $80\%$. In practice,
these settings of hash table work well and significantly outperform
red-black tree. All the numerical results in this paper are produced
with hash table.

Besides the expensive data accessing step, the computational
expensive step is the evaluation of $H_{:,i^\itellp}$.  Let $N_H =
\max_i \abs{\hset{i}}$ be the maximum number of nonzero entries in
columns of the Hamiltonian matrix.  Although $N_H \ll N_{\text{FCI}}$,
$N_H$ still scales as $\bigO(\nelec^2 \norb^2)$.  The computational
cost for evaluating each entry $H_{i,j}$ also depends moderately on
$\nelec$. \sname{} uses an efficient Fortran implemented open source
quantum chemistry code \textsc{Hande-QMC} as backend for the evaluation
of Hamiltonian entries.

Shared memory parallelism based on OpenMP is used in our
implementation. For each iteration, the double excitation calculation
is the bottleneck for the evaluation $H_{:,i^\itellp}$, which is
embarrassingly parallelized with OpenMP.  In terms of runtime, accessing
a nonzero coefficient of $\vec{b}$ and $\vec{c}$ is also expensive due
to the lack of memory continuity.  Therefore, we also parallelize the
access to $b_{\hset{i^\itellp}}$ and $c_{\hset{i^\itellp}}$
with OpenMP. Due to the possible collision of the hash function of open
addressing, we partition the hash array into $2000$ blocks and set
locker for each block, such that no two threads can access the same
block simultaneously. Increasing the number of lockers would reduce
the idling time of threads but would increase the memory cost. We did
not try to optimize the number of blocks.

Last point on implementation focuses on the numerical stability of the
Rayleigh quotient. Different from other iterative methods, \sname{}
updates one determinant per iteration. Hence, for large systems,
the number of iteration could easily go beyond $10^8$. For cumulated
quantities such as $\vec{c}^\top \vec{c}$ and $\vec{c}^\top H \vec{c}$,
the value is updated at least $10^8$ times, hence the accumulated
numerical error could pollute the chemical accuracy, and thus careful
treatment is needed.  In our implementation, we use a quadruple-precision
floating point for $\vec{c}^\top \vec{c}$ and $\vec{c}^\top H \vec{c}$
such that the relative error is at most $10^{-16}$ unless the number of
iteration exceeds $10^{16}$.

Let us remark that our current implementation of \sname{} is by no
means optimal. The bottleneck of the current implementation is the
na\"{i}ve hash table. Random access to the main memory is expensive
since the cache hierarchy is not fully adapted. An optimized hash table
may improve the performance by a big constant.

\smallskip

To conclude the section, let us conduct a leading order per iteration
complexity analysis for \sname{}.  In determinant-select step,
since all $b_i^\itell$ and $c_i^\itell$ for $i \in
\hset{i^\itellp}$ have been accessed in the previous iteration,
$i^\itellp$ can be computed without paying the cost of accessing the
data structure of $\vec{b}$ and $\vec{c}$. Hence the leading cost is
$\bigO(N_H)$ with a small prefactor. Line search and updating $\vec{c}$
cost $\bigO(1)$ operations and are hence negligible. Updating $\vec{b}$
is the most expensive step throughout the algorithm.  It requires
evaluating $\bigO(N_H)$ entries of Hamiltonian matrix and
accessing $\vec{b}$ and $\vec{c}$ $\bigO(N_H)$ times. This step costs
$\bigO(N_H)$ operation with a prefactor being the Hamiltonian per entry
evaluation cost plus the averaged data structure accessing cost. In our
implementation, the compression step is combined with the updating step.
Once $H_{:,i^\itellp}$ has been evaluated, $b_i^\itell$ and
$c_i^\itell$ for $i \in \hset{i^\itellp}$ are accessed,
then exact update of $b_{i^\itellp}^\itellp$, cumulative
updates of $\vec{c}^\top \vec{c}$ and $\vec{c}^\top H \vec{c}$ cost
$\bigO(N_H)$ operations with a small prefactor. Overall, \sname{} costs
$\bigO(N_H)$ operations per iteration with the prefactor dominated
by the computation cost of one Hamiltonian entry and the averaged
access cost of the data structure.  The memory cost of \sname{}
is dominated by the cost of allocating the data structure.

\section{Numerical results}\label{sec:numres}

In this section, we perform a sequence of numerical experiments
to demonstrate the efficiency of \sname{}. First, we compare the
performance of \sname{}, Heat-bath CI (HCI), DMRG and \fciqmc{} (FCIQMC
with initiator and semi-stochastic adaptation) on \ch{H2O}, \ch{C2}
and \ch{N2} under cc-pVDZ basis. Then, we benchmark the binding curve
of nitrogen dimer under cc-pVDZ basis using \sname{} up to $10^{-3}$
mHa accuracy. Finally, we use \sname{} to calculate the ground state
energy of chromium dimer \ch{Cr2} under the Ahlrichs VDZ Basis at $r =
1.5$\r{A}, which is a well-known challenging task due to the strong
correlation.

In all experiments, the orbitals and integrals are calculated via
restricted Hartree Fock (RHF) in \textsc{Psi4}\cite{Parrish2017}
package. All the reported energies are in Hartree (Ha) but the length
unit is in either Bohr radius ($a_0$) or \r{a}ngstr\"{o}m (\r{A})
due to different configurations in the references.

\subsection{Numerical results of \ch{H2O}, \ch{C2}, and \ch{N2}}
\label{sec:num-res-molecule}

% Hardware and Software environment
We first compare the performance of \sname{} with other algorithms,
HCI, DMRG and \fciqmc{}.  In this paper, we choose \fciqmc{} instead
of FCIQMC or iFCIQMC because it balances well among bias, variance and
runtime.  \sname{} is implemented as stated in Section \ref{sec:impl}
with the on-the-fly Hamiltonian elements evaluation interfaced from
\textsc{Hande-QMC}\cite{Spencer2015,Spencer2018}.  HCI adopts the
original implementation in \textsc{Dice}~\cite{Holmes2016}; DMRG
adopts the widely used implementation in \textsc{Block}\cite{Chan2002,
Chan2004, Ghosh2008, Sharma2012, Olivares-Amaya2015}; \fciqmc{} adopts
the implementation in \textsc{NECI} code~\cite{Booth2014}. All programs
are compiled by Intel compiler 19.0.144 with \texttt{-O3} option. MPI
and OpenMP support are disabled for all programs in this section. All
the tests in this section are produced on a machine with Intel Xeon
CPU E5-1650 v2 @ 3.50GHz and $64$GB memory.
% Algorithms Parameters
For all algorithms, the Hartree-Fock state is used as the initial
wavefunction. Most parameters in \sname{}, HCI, DMRG, and \fciqmc{}
will be clearly stated in the later content. We run HCI with different
truncation threshold $\varepsilon_1$, DMRG with different maximum
bond dimension $\max M$ and \fciqmc{} with different target population
$m$. Any unspecified parameter is set to be the default value. Besides
the specified or default parameters, for \fciqmc{}, the time step is
optimized by \textsc{NECI}; the initiator truncation threshold is $3$ and
the size of the deterministic space is 1000. For all the tests reported
in Table~\ref{tab:h2o}, Table~\ref{tab:c2}, and Table~\ref{tab:n2},
\fciqmc{} runs for $10000$ steps and the energy is estimated by the
block analysis in \textsc{NECI} . In all algorithms, the energy is
reported without any perturbation or extrapolation post-calculation.
Variational energy (Rayleigh quotient) is reported for \sname{}, HCI and
DMRG, while average projected energy is reported for \fciqmc{}. Since
\fciqmc{} is a stochastic method, we also report one significant digit
of the Standard Error of the Mean (SEM) in the parenthesis and the SEM
is of the same accuracy level as the last digit of the average. SEM is
estimated as the sample standard deviation divided by the square root
of the sample size. Also Root Mean Square Error (RMSE) of the average
estimator is adopted as the error of \fciqmc{}, which is defined as
$\sqrt{\text{standard error}^2 + \mathrm{bias}^2}$. We emphasize that
similar perturbation calculation as in HCI can also be applied to
\sname{} and the comparison is left as future work.

% Systems Specification
In this section, we test the four algorithms on three molecules:
\ch{H2O} with OH bond length 1.84345$a_0$ and HOH bond angle
110.6\textdegree~\cite{Olsen1998, Booth2009}, \ch{C2} with bond length
1.24253\r{A}~\cite{Holmes2016, Sharma2015} and \ch{N2} with bond
length 2.118$a_0$~\cite{Chan2004b}. The properties of the systems are
summarized in Table~\ref{tab:system}, where the reference ground state
energy is calculated by \sname{} to a high precision.

\begin{table*}[htb]
    \centering
    \begin{tabular}{ccccccc}
    \toprule
    Molecules & Basis & Electrons & Orbitals & Dimension & HF Energy &
    GS Energy \\
    \midrule
    \ch{H2O} & cc-pVDZ & 10 & 24 & $4.53\times 10^8$ &
    -76.0240386 & -76.2418601  \\
    \ch{C2}  & cc-pVDZ & 12 & 28 & $1.77\times 10^{10}$ &
    -75.4168820 & -75.7319604  \\
    \ch{N2}  & cc-pVDZ & 14 & 28 & $1.75\times 10^{11}$ &
    -108.9493779 & -109.2821727 \\
    \bottomrule
    \end{tabular}
    \caption{Properties of test molecule systems. HF energy and GS energy
    denote Hartree-Fock energy and ground state energy respectively.}
    \label{tab:system}
\end{table*}

\subsubsection{\ch{H2O} molecule}
\label{sec:h2o}

\begin{figure}[ht]
    \centering
    \includegraphics[width=0.5\textwidth]{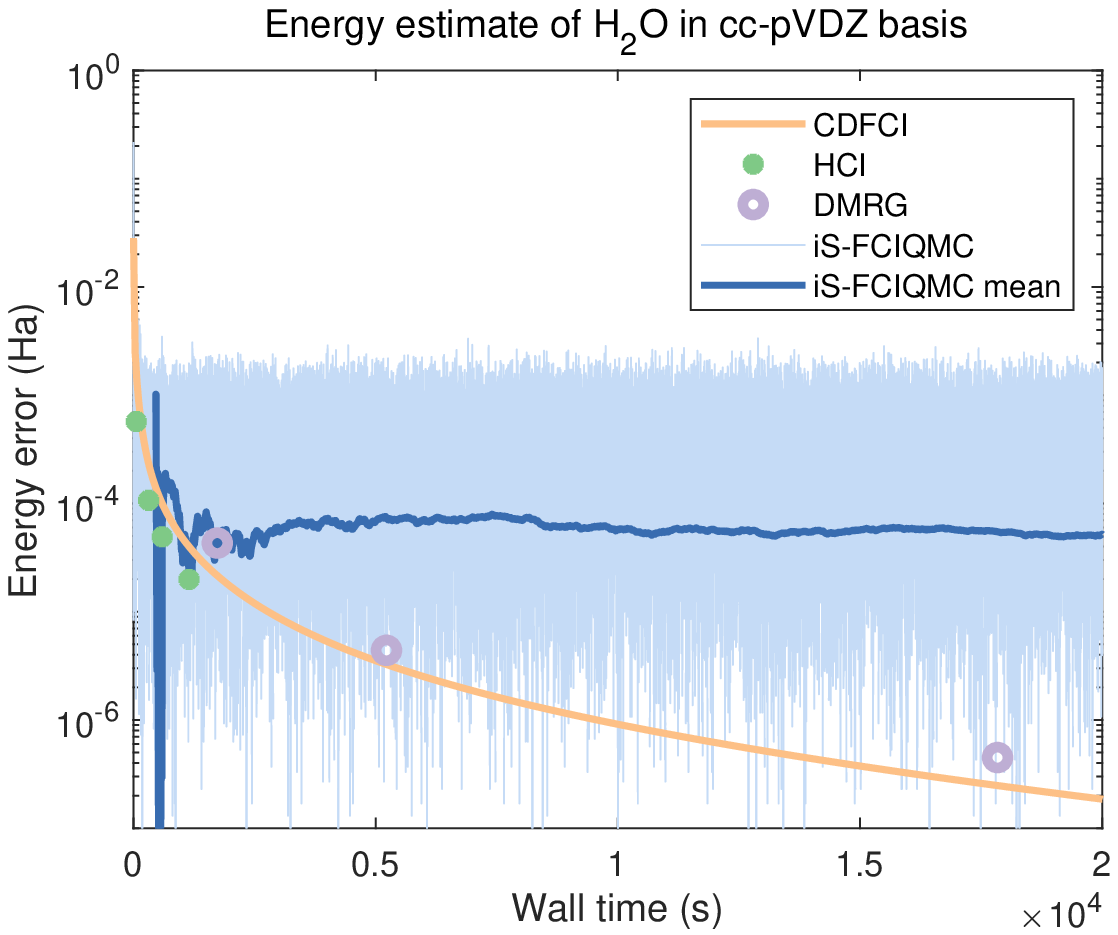}
    \caption{Convergence of ground state energy of \ch{H2O} against
    wall clock time. Each point or curve represents one test as in
    Table~\ref{tab:h2o}. For \fciqmc{}, the projected energy and its
    cumulative average from iteration $5000$ are plotted with target
    population $m=50000$. } \label{fig:h2o}
\end{figure}

Table~\ref{tab:h2o} and Figure~\ref{fig:h2o} illustrate numerical results
for \ch{H2O}. In Table~\ref{tab:h2o}, we report the detailed results
and the corresponding used parameters.  Figure~\ref{fig:h2o} plots
the convergence of the energy against the wall-clock time based on the
results in Table~\ref{tab:h2o}. For \fciqmc{}, we run another test with
$m = 50000$ for longer time and plot the curve of the projected energy
as well as the cumulative average of energy starting at iteration $5000$.

\begin{table*}[ht]
    \centering
    \begin{tabular}{cl|lcr|rr}
        \toprule
        \multirow{2}{*}{Algorithm} & \multirow{2}{*}{Parameter}
        & \multirow{2}{*}{Energy} & \multirow{2}{*}{Error}
        & \multirow{2}{*}{Time(s)} & \multicolumn{2}{c}{\sname{}} \\
        & & & & & Time(s) & Ratio \\
        \midrule
        \multirow{7}{*}{\sname{}} & \multirow{7}{*}{$\varepsilon = 0$}
        &   -76.2318601 & $1.0 \times 10^{-2}$ & 3.7     & - \\
        & & -76.2408601 & $1.0 \times 10^{-3}$ & 96.2    & - \\
        & & -76.2417601 & $1.0 \times 10^{-4}$ & 592.5   & - \\
        & & -76.2418501 & $1.0 \times 10^{-5}$ & 2780.0  & - \\
        & & -76.2418591 & $1.0 \times 10^{-6}$ & 9569.5  & - \\
        & & -76.2418600 & $1.0 \times 10^{-7}$ & 25227.5 & - \\
        & & -76.2418601 & $1.0 \times 10^{-8}$ & 54242.2 & - \\
        \midrule
        \multirow{4}{*}{HCI}
        & $\varepsilon_1 = 1.0\times 10^{-4}$ & -76.2412891
        & $5.7\times 10^{-4}$ & 58.4   & 156.3  & 0.37x \\
        & $\varepsilon_1 = 2.0\times 10^{-5}$ & -76.2417533
        & $1.1\times 10^{-4}$ & 312.9  & 565.3  & 0.55x \\
        & $\varepsilon_1 = 1.0\times 10^{-5}$ & -76.2418109
        & $4.9\times 10^{-5}$ & 593.5  & 993.1  & 0.60x \\
        & $\varepsilon_1 = 5.0\times 10^{-6}$ & -76.2418402
        & $2.0\times 10^{-5}$ & 1148.3 & 1823.2 & 0.63x \\
        \midrule
        \multirow{4}{*}{DMRG}
        &  $\max M = 500$ & -76.2418170 & $4.3\times 10^{-5}$
        &  1731 & 1089.7 & 1.59x \\
        & $\max M = 1000$ & -76.2418557 & $4.4\times 10^{-6}$
        &  5224 & 4435.7 & 1.18x \\
        & $\max M = 2000$ & -76.2418596 & $4.5\times 10^{-7}$
        & 17839 & 13802.6 & 1.29x \\
        & $\max M = 4000$ & -76.2418599 & $1.7\times 10^{-7}$
        & 77023 & 20585.9 & 3.74x \\
        \midrule
        \multirow{4}{*}{\fciqmc{}}
        &   $m = 10000$ & -76.2418(3) & $ 2.7\times 10^{-4}$
        & 222.4  & 277.2  & 0.80x \\
        &   $m = 50000$ & -76.24197(8) & $1.4\times 10^{-4}$
        & 1009.0 & 470.5  & 2.14x \\
        &  $m = 100000$ & -76.24181(5) & $7.1\times 10^{-5}$
        & 1942.8 & 762.1  & 2.55x \\
        &  $m = 500000$ & -76.24181(3) & $5.9\times 10^{-5}$
        & 9074.3 & 875.2  & 10.37x \\
        \bottomrule
    \end{tabular}
    \caption{Convergence of ground state energy of \ch{H2O}. For
    \sname{}, we run the test once and report the wall time to reach
    the accuracy. For other tests, each row corresponds to one test.
    We also report the wall time for \sname{} to reach the same accuracy
    as well as the ratio of the wall time of the method over the wall
    time of \sname{} in the last two columns.  \fciqmc{} runs for
    $10000$ iterations and reports the average projected energy and the
    standard error (in parenthesis) through block analysis in the energy
    column. RMSE is reported as the error of \fciqmc{}.} \label{tab:h2o}
\end{table*}

From Table~\ref{tab:h2o} and Figure~\ref{fig:h2o}, we shall see that
all algorithms reach chemical accuracy efficiently. \sname{} has a
good performance in general. The energy drops quickly to the level of
$0.1$ mHa accuracy at the beginning. Then it has a slower but steady
linear decay. This behavior proves the rationale behind \sname{}:
contributions of different determinants to the FCI energy vary a lot,
especially in the early stage of iterations. Since \sname{} always
updates the ``best'' determinant at each iteration, it is able to
reach high accuracy with only a few iterations.

For small molecules like \ch{H2O}, HCI is also an excellent
algorithm and costs less time to achieve the same accuracy comparing
to \sname{}. It shows that the determinant selecting strategy used in
HCI, which relies on decaying property of Hamiltonian entries, is also
quite efficient for molecules.  For example, when $\varepsilon_1 =
5.0 \times 10^{-6}$, HCI uses only $3006594$ determinants to reach
$2.0\times 10^{-5}$ Ha accuracy, whereas \sname{} uses $1823176$
determinants, which is about $60\%$ determinants used by HCI to achieve
the same accuracy.

The speedup of HCI over \sname{} is due to the different
implementation strategies of the algorithms.  The implementation of
HCI in \textsc{Dice} stores the submatrix of the Hamiltonian with
respect to the selected determinants in the main memory, and reuses
them for inner Davidson iterations. Both the submatrix and the vector
are stored and accessed in contiguous memory. Therefore, two
advantages of the implementation come into play: one-time evaluation
of Hamiltonian entries and efficient usage of memory
hierarchy. However, the disadvantage is also obvious: huge memory cost
for the submatrix. In Table~\ref{tab:h2o} and Figure~\ref{fig:h2o}, we
do not report results for smaller $\varepsilon_1$ because
\textsc{Dice} reaches the memory limit. The high memory cost is also
the reason why the variational stage of HCI does not perform good for
chromium dimer (see Section~\ref{sec:cr2}). As a comparison, \sname{}
uses a different philosophy in the implementation. \sname{} calculates
the Hamiltonian entries on-the-fly, which saves all memory for the
coefficient vector, and stores the coefficients in a hash table. Hence
much more coefficients can be used to represent the ground state but
paying the cost of repeated evaluation of Hamiltonian entries and
limited usage of memory hierarchy.

DMRG also achieves high accuracy in reasonable time with small memory
cost. But it is always slower than \sname{} and HCI for \ch{H2O}.

\fciqmc{} is very efficient for small number of walkers and
iterations. It is able to achieve reasonable accuracy in a short
time. In Figure~\ref{fig:h2o} we see the convergence behavior of
\fciqmc{} projected energy. It can reach accuracy level of $1$ mHa very
efficiently, but hard to converge to higher accuracy due to the slow
convergence of Monte Carlo and the bias introduced by the initiator
approximation. It could be possible to use more walkers to reduce the
variance and bias. However, as shown in Table~\ref{tab:h2o}, moderate
increase of the amount of walkers does not change the convergence
behavior.

\subsubsection{Carbon dimer and nitrogen dimer}
\label{sec:c2n2}

\begin{table*}[ht]
    \centering
    \begin{tabular}{cl|lcr|rr}
        \toprule
        \multirow{2}{*}{Algorithm} & \multirow{2}{*}{Parameter}
        & \multirow{2}{*}{Energy} & \multirow{2}{*}{Error}
        & \multirow{2}{*}{Time(s)} & \multicolumn{2}{c}{\sname{}} \\
        & & & & & Time(s) & Ratio \\
        \midrule
        \multirow{5}{*}{\sname{}}
        & \multirow{5}{*}{$\varepsilon = 3.0 \times 10^{-8}$}
          & -75.7219604 & $1.0 \times 10^{-2}$ & $49.0$    & - & \\
        & & -75.7309604 & $1.0 \times 10^{-3}$ & $388.2$   & - & \\
        & & -75.7318604 & $1.0 \times 10^{-4}$ & $2687.3$  & - & \\
        & & -75.7319504 & $1.0 \times 10^{-5}$ & $13717.6$ & - & \\
        & & -75.7319594 & $1.0 \times 10^{-6}$ & $55210.2$ & - & \\
        \midrule
        \multirow{4}{*}{HCI}
        & $\varepsilon_1 = 1.0\times 10^{-4}$ & -75.7305361
        & $1.4\times 10^{-3}$ &  100.9 &  277.8 & 0.36x \\
        & $\varepsilon_1 = 2.0\times 10^{-5}$ & -75.7317130
        & $2.5\times 10^{-4}$ &  745.0 & 1319.1 & 0.56x \\
        & $\varepsilon_1 = 1.0\times 10^{-5}$ & -75.7318541
        & $1.1\times 10^{-4}$ & 1261.8 & 2565.2 & 0.49x \\
        & $\varepsilon_1 = 5.0\times 10^{-6}$ & -75.7319170
        & $4.4\times 10^{-5}$ & 2644.3 & 4989.8 & 0.53x \\
        \midrule
        \multirow{4}{*}{DMRG}
        & $\max M = 500$  & -75.7312704 & $6.9\times 10^{-4}$
        &  8624 &  544.4 & 15.84x \\
        & $\max M = 1000$ & -75.7318227 & $1.4\times 10^{-4}$
        & 14163 & 2102.9 & 6.73x \\
        & $\max M = 2000$ & -75.7319403 & $2.0\times 10^{-5}$
        & 24377 & 8582.9 & 2.84x \\
        & $\max M = 4000$ & -75.7319583 & $2.2\times 10^{-6}$
        & 68071 & 35435  & 1.92x \\
        \midrule
        \multirow{5}{*}{\fciqmc{}}
        &   $m = 10000$ & -75.729(1)   & $ 2.9\times 10^{-3}$
        & 229.5  & 140.8 & 1.63x \\
        &   $m = 50000$ & -75.7301(5) & $ 1.9\times 10^{-3}$
        & 1041.4 & 212.9 & 4.89x \\
        &  $m = 100000$ & -75.7314(4) & $ 7.0\times 10^{-4}$
        & 2038.2 & 539.8 & 3.78x \\
        &  $m = 500000$ & -75.7320(1) & $ 1.5\times 10^{-4}$
        & 9604.2 & 2003.2 & 4.79x \\
        & $m = 1000000$ & -75.7318(1) & $ 2.1\times 10^{-4}$
        & 18644.8 & 1497.4 & 12.45x \\
        \bottomrule
    \end{tabular}
    \caption{Convergence of ground state energy of \ch{C2}.  \fciqmc{}
    runs for $10000$ iterations and reports the average projected energy
    and standard error (in parenthesis) through the block analysis in
    the energy column. RMSE is reported as the error of \fciqmc{}.}
    \label{tab:c2}
\end{table*}

\begin{figure}[ht]
    \centering \includegraphics[width=0.5\textwidth]{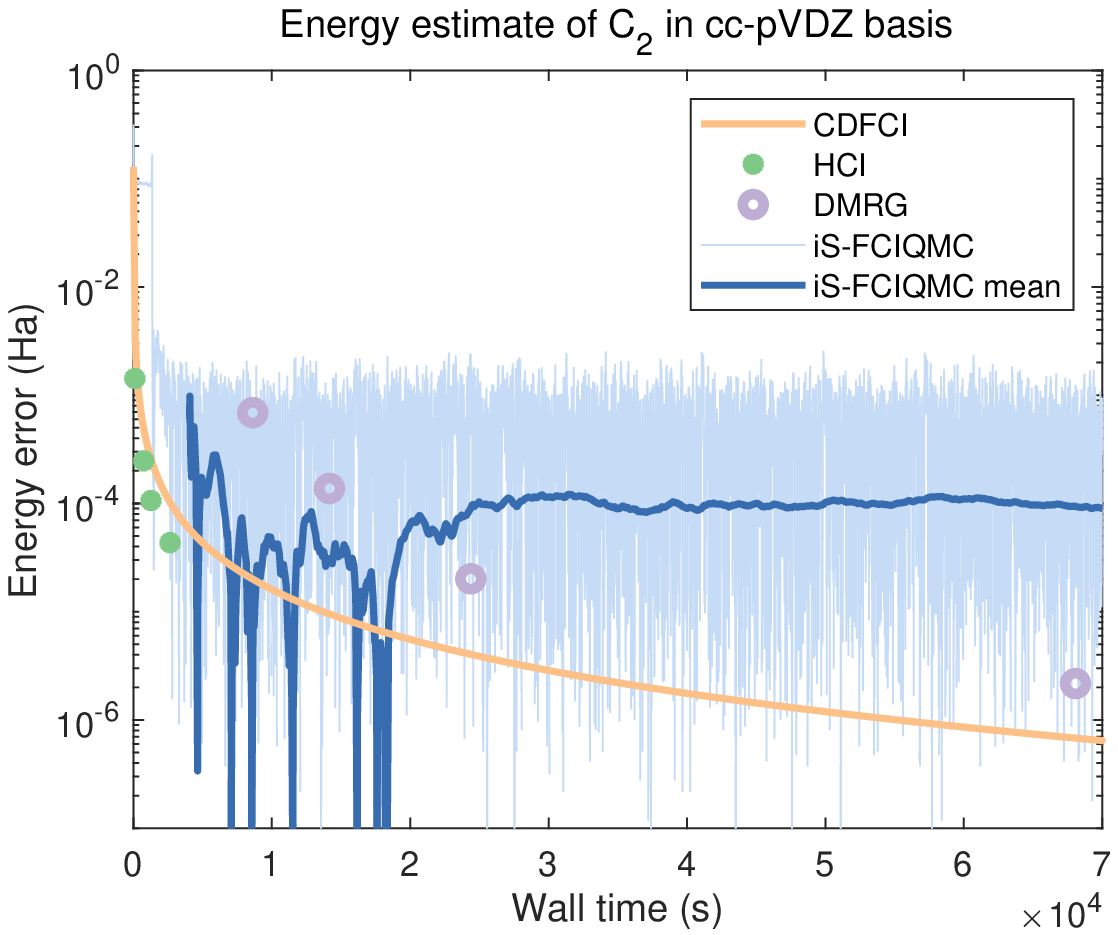}
    \caption{Convergence of ground state energy of \ch{C2} against
    wall clock time. Each point or curve represents one test as in
    Table~\ref{tab:c2}. For \fciqmc{}, the projected energy and its
    cumulative average from iteration $5000$ are plotted with target
    population $m=500000$.} \label{fig:c2}
\end{figure}

% N2
\begin{table*}[ht]
    \centering
    \begin{tabular}{cl|lcr|rr}
        \toprule
        \multirow{2}{*}{Algorithm} & \multirow{2}{*}{Parameter}
        & \multirow{2}{*}{Energy} & \multirow{2}{*}{Error}
        & \multirow{2}{*}{Time(s)} & \multicolumn{2}{c}{\sname{}} \\
        & & & & & Time(s) & Ratio \\
        \midrule
        \multirow{5}{*}{\sname{}}
        & \multirow{5}{*}{$\varepsilon = 5.0\times 10^{-7}$}
          & -109.2721727 & $1.0\times 10^{-2}$ & $33.4$    & - & \\
        & & -109.2811727 & $1.0\times 10^{-3}$ & $752.6$   & - & \\
        & & -109.2820727 & $1.0\times 10^{-4}$ & $7892.6$  & - & \\
        & & -109.2821627 & $1.0\times 10^{-5}$ & $49862.6$ & - & \\
        \midrule
        \multirow{4}{*}{HCI}
        & $\varepsilon_1 = 1.0\times 10^{-4}$ & -109.2805259
        & $1.7\times 10^{-3}$ &  100.7 &  427.1 & 0.24x \\
        & $\varepsilon_1 = 2.0\times 10^{-5}$ & -109.2817822
        & $3.9\times 10^{-4}$ &  730.9 & 2107.4 & 0.35x \\
        & $\varepsilon_1 = 1.0\times 10^{-5}$ & -109.2819787
        & $1.9\times 10^{-4}$ & 1335.2 & 4266.8 & 0.31x \\
        & $\varepsilon_1 = 5.0\times 10^{-6}$ & -109.2820857
        & $8.7\times 10^{-5}$ & 3330.0 & 8920.3 & 0.37x \\
        \midrule
        \multirow{4}{*}{DMRG}
        &  $\max M = 500$ & -109.2809830 & $1.2\times 10^{-3}$
        &  9936 &   619.6 & 16.04x \\ 
        & $\max M = 1000$ & -109.2818757 & $3.0\times 10^{-4}$
        & 17647 &  2806.7 & 6.29x \\
        & $\max M = 2000$ & -109.2821098 & $6.3\times 10^{-5}$
        & 37549 & 11857.1 & 3.12x \\
        & $\max M = 4000$ & -109.2821632 & $9.5\times 10^{-6}$
        & 85703 & 51574.7 & 1.66x \\
        \midrule
        \multirow{5}{*}{\fciqmc{}}
        &   $m = 10000$ & -109.2818(4) & $ 5.2\times 10^{-4}$
        & 235.7  & 1556.8  & 0.15x \\
        &   $m = 50000$ & -109.2822(3) & $ 2.6\times 10^{-4}$
        & 1068.9 & 3161.5 & 0.34x \\
        &  $m = 100000$ & -109.2818(2) & $ 4.0\times 10^{-4}$
        & 2090.5 & 2077.2  & 1.01x \\
        &  $m = 500000$ & -109.2822(1) & $ 1.2\times 10^{-4}$
        & 9839.3 & 6832.6 & 1.44x \\
        & $m = 1000000$ & -109.28214(5) & $ 5.0\times 10^{-5}$
        & 18959.7 & 14510.0 & 1.31x \\
        \bottomrule
    \end{tabular}
    \caption{Convergence of energy of \ch{N2}.  \fciqmc{} runs for
    $10000$ iterations and reports the average of projected energy and
    standard error (in parenthesis) through block analysis. RMSE is
    reported as the error of \fciqmc{}.} \label{tab:n2}
\end{table*}

\begin{figure}[ht]
    \centering
    \includegraphics[width=0.5\textwidth]{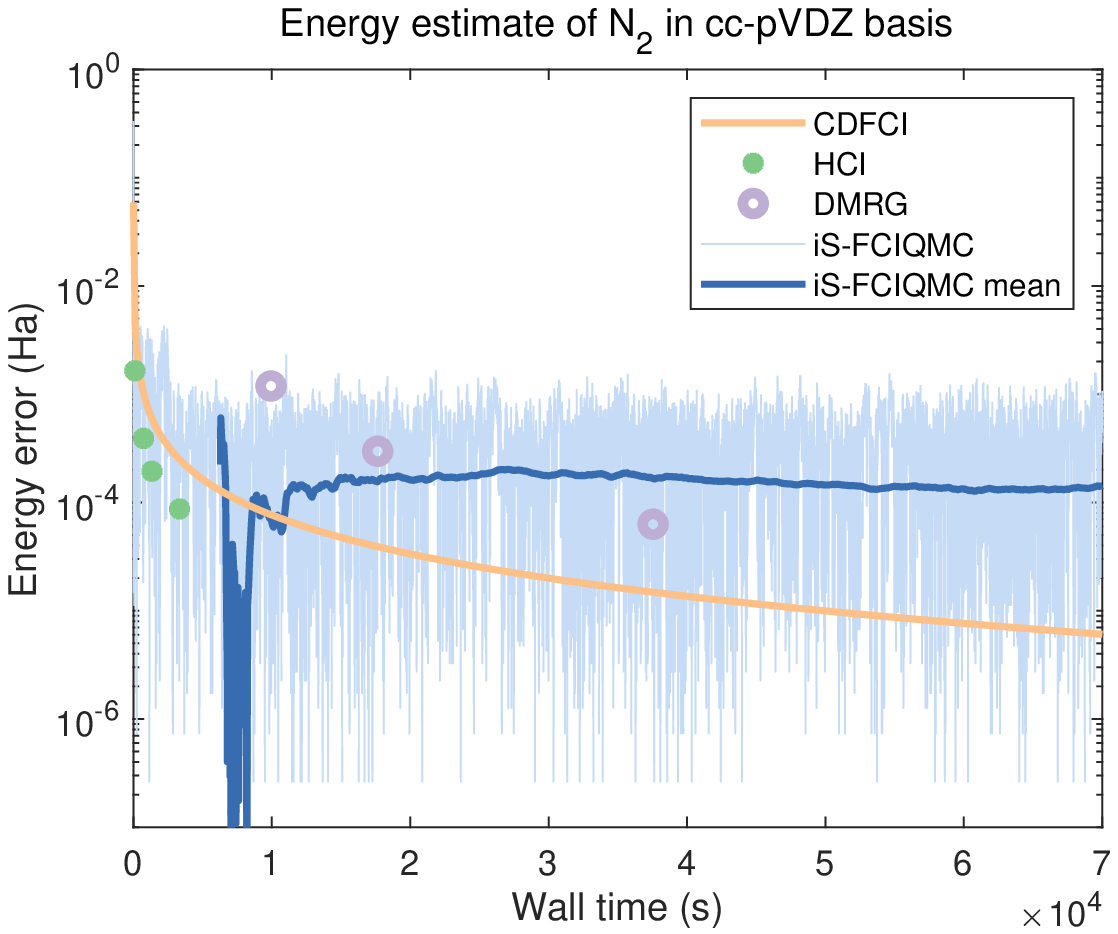}
    \caption{Convergence of ground state energy of \ch{N2} against
    wall clock time. Each point or curve represents one test as in
    Table~\ref{tab:n2}.  For \fciqmc{}, the projected energy and its
    cumulative average from iteration $5000$ are plotted with target
    population $m=500000$.} \label{fig:n2}
\end{figure}

Carbon dimer \ch{C2} and nitrogen dimer \ch{N2} are more challenging
than \ch{H2O} molecule since their correlation is stronger and
the dimension $N_{\text{FCI}}$ is higher. The results of \ch{C2}
are reported in Table~\ref{tab:c2} and Figure~\ref{fig:c2};
and the results of \ch{N2} are reported in Table~\ref{tab:n2} and
Figure~\ref{fig:n2}. In general, \ch{C2} costs more time than \ch{H2O}
to converge to a fixed accuracy and \ch{N2} costs more time than
\ch{C2}, which agrees with their system complexities.

\sname{} shows similar convergence pattern for \ch{C2} and \ch{N2},
with fast decay at the beginning followed by a slower but steady
linear decay. It takes only several minutes to reach the chemical
accuracy. Therefore, \sname{} is consistently efficient for different
systems with different correlation strength.

HCI also shows similar convergence behavior for \ch{C2} and \ch{N2}. It
converges to chemical accuracy the fastest among tested algorithms.
However, HCI can not converge to higher accuracy due to the memory
limit of the implementation. Comparing to \sname{} in terms of the
number of operations, however, \sname{} uses less operations and
determinants than HCI to the same accuracy level, as in the case
of \ch{H2O}.

DMRG also performs similar for \ch{C2} and \ch{N2} but is significantly
slower than \sname{} and HCI. One reason is that DMRG needs more
iterations to converge due to the strong correlation. In this sense,
determinant selecting algorithms are less affected by the correlation
strength than DMRG.

\fciqmc{}, however, has a different behavior for \ch{C2} and
\ch{N2}. While \ch{N2} is significantly harder than \ch{C2} for \sname{},
HCI and DMRG, \fciqmc{} reaches a higher accuracy for \ch{N2} within
$10000$ iterations, as shown by the data in Table~\ref{tab:c2} and
Table~\ref{tab:n2}.  We point out that \fciqmc{} performs quite well
for \ch{N2}, as it can reach $10^{-4}$ Ha error in a short time with
only $m=10000$ or $m=50000$ walkers, whereas \sname{} takes more time
to converge since the dimension of \ch{N2} is higher than \ch{H2O}
and \ch{C2}. \fciqmc{} seems to be less influenced by the increase
of dimensionality.

In conclusion, as shown in both Section~\ref{sec:h2o} and
Section~\ref{sec:c2n2}, \sname{} is efficient for both weakly-correlated
and strongly-correlated systems. It can achieve chemical accuracy
efficiently and is able to achieve higher accuracy in all tested
molecules. HCI costs more operations than \sname{} to achieve the
same accuracy but costs less time due to the different philosophies in
implementations. Comparing to \sname{} and HCI, DMRG is less efficient
for strongly-correlated systems. While similar as \sname{}, DMRG can
also achieve much higher accuracy than the chemical accuracy.  \fciqmc{}
is also efficient to reach chemical accuracy with a few walkers in
short time for the testing molecules, but it may need much more time
and walkers to reach higher accuracy.

\begin{figure}[ht]
    \centering
    \includegraphics[width=0.5\textwidth]{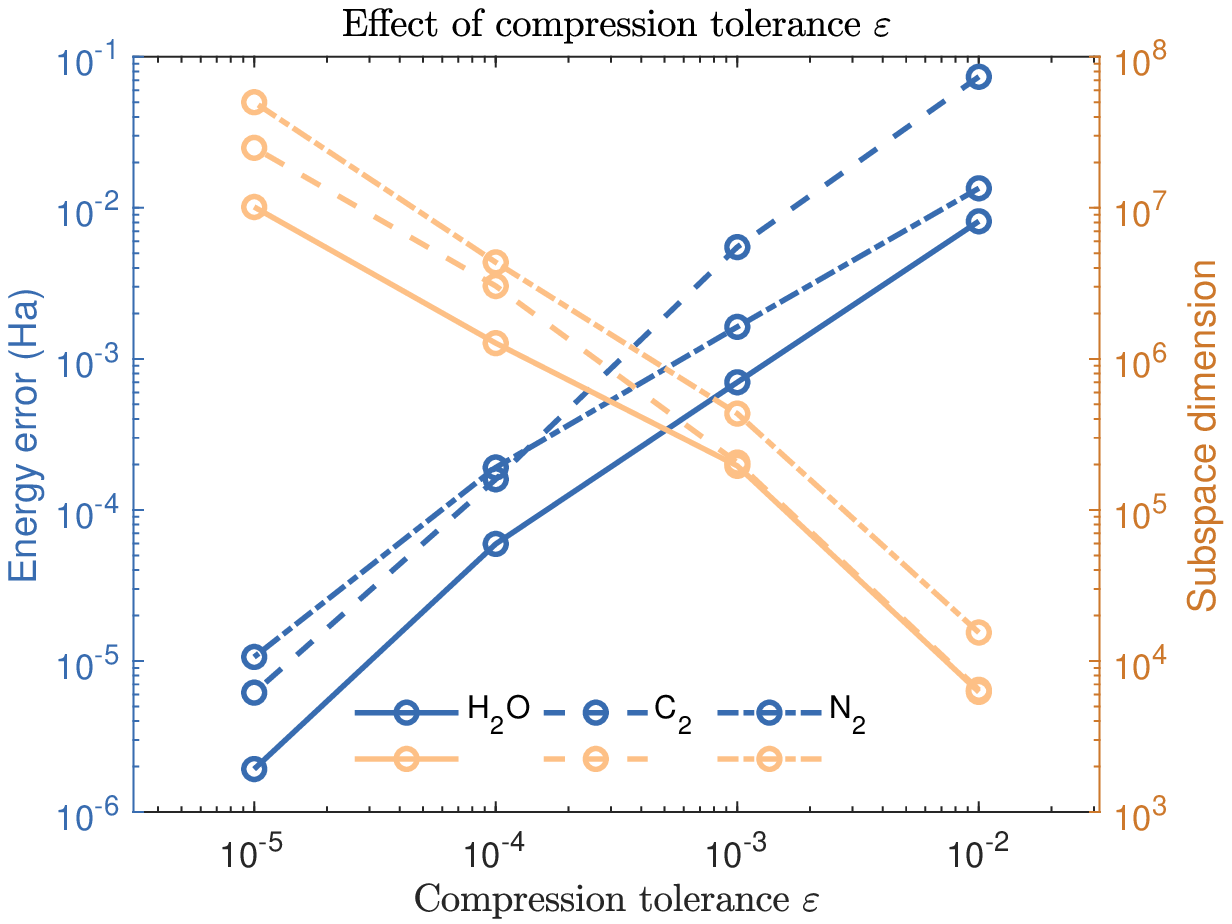}
    \caption{The effect of compression tolerance $\varepsilon$ in
    \sname{}. The figure shows the error of energy and the number of
    nonzeros entries in the vector $\vec{b}$ after convergence for
    differenct $\varepsilon$. By choosing differenct $\varepsilon$,
    we can easily balance between accuracy and memory cost.}
    \label{fig:eps}
\end{figure}

In terms of the usability, \sname{}, HCI, and DMRG only have one single
parameter to be tuned, whereas \fciqmc{} has more parameters. The proper
parameter in \sname{} can be revealed in a few minutes, judging from
whether the stabilized vector $\vec{b}$ properly utilizes the given
amount of memory. By choosing $\varepsilon$, we can easily balance
between accuracy and memory cost, as shown in Figure~\ref{fig:eps}. We
conclude that \sname{} is an easy-to-use efficient algorithm for FCI
problems.

\subsection{Binding curve of \ch{N2}}
\label{sec:bcn2}

In this section, we benchmark the all-electron nitrogen binding curve
using \sname{} under the Dunning's cc-pVDZ basis. The nitrogen binding
curve is a well-known difficult problem. When the nitrogen atoms are
stretched away from the equilibrium geometry, Hartree-Fock theory no
longer gives a good approximation and the system becomes multi-referenced
due to the triple bond between the atoms. DMRG and coupled cluster
theory, e.g., CCSD, CCSD(T), CCSDT, etc., have been tested on this
problem, but only on 6 geometry configurations~\cite{Chan2004b}.
Here we show that \sname{} is capable to efficiently benchmark the
all-electron nitrogen binding curve on a very fine grid of bond length
and the variational energy converges to at least $10^{-3}$ mHa accuracy
in each configuration.

\begin{figure}[ht]
    \centering
    \subfigure[Energy of \ch{N2}]{
    \includegraphics[width=0.5\textwidth]{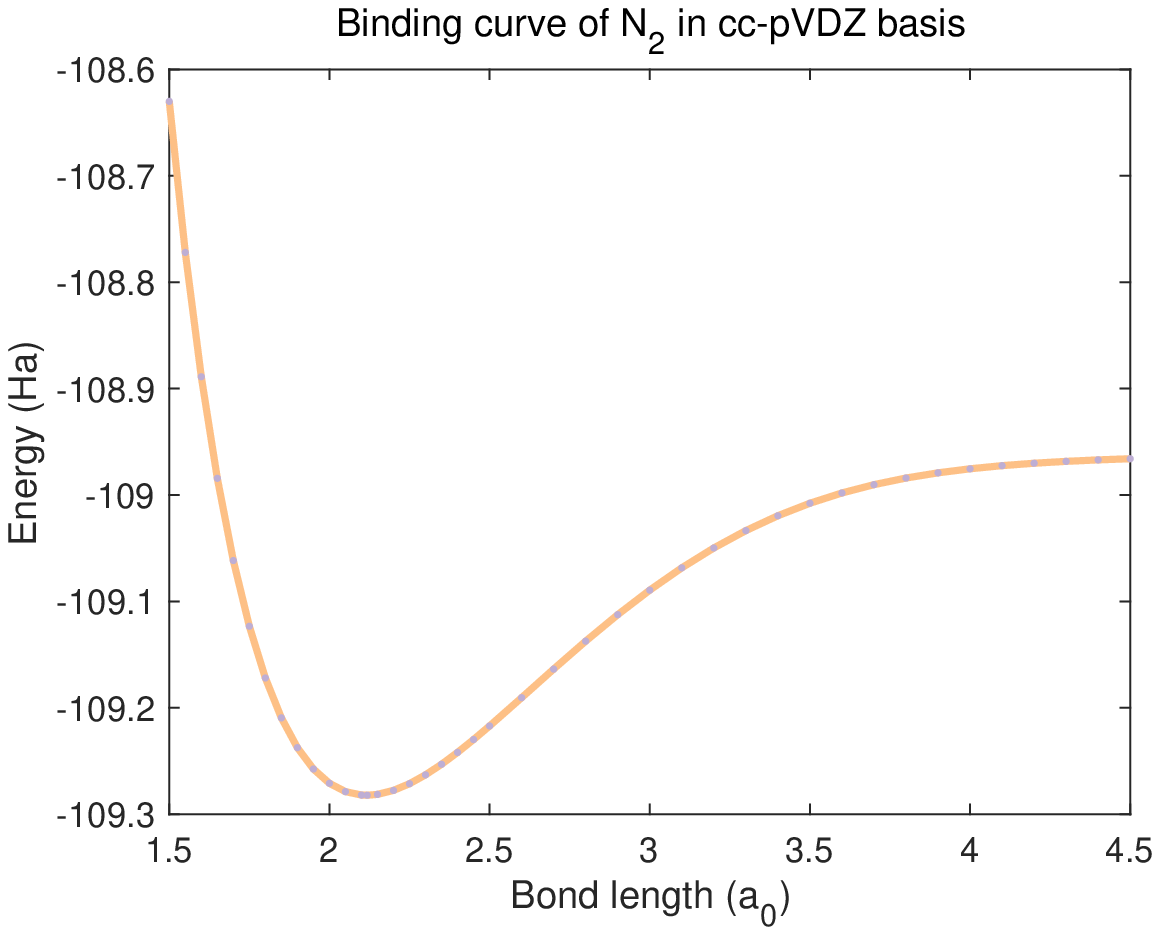}
    }
    \subfigure[Hartree-Fock value
    $\vec{c}_{\mathrm{HF}}/\norm{\vec{c}}$ and the number of nonzeros
    of $\vec{c}$ to reach $10^{-5}$ Ha energy error for \sname{}]{
    \includegraphics[width=0.5\textwidth]{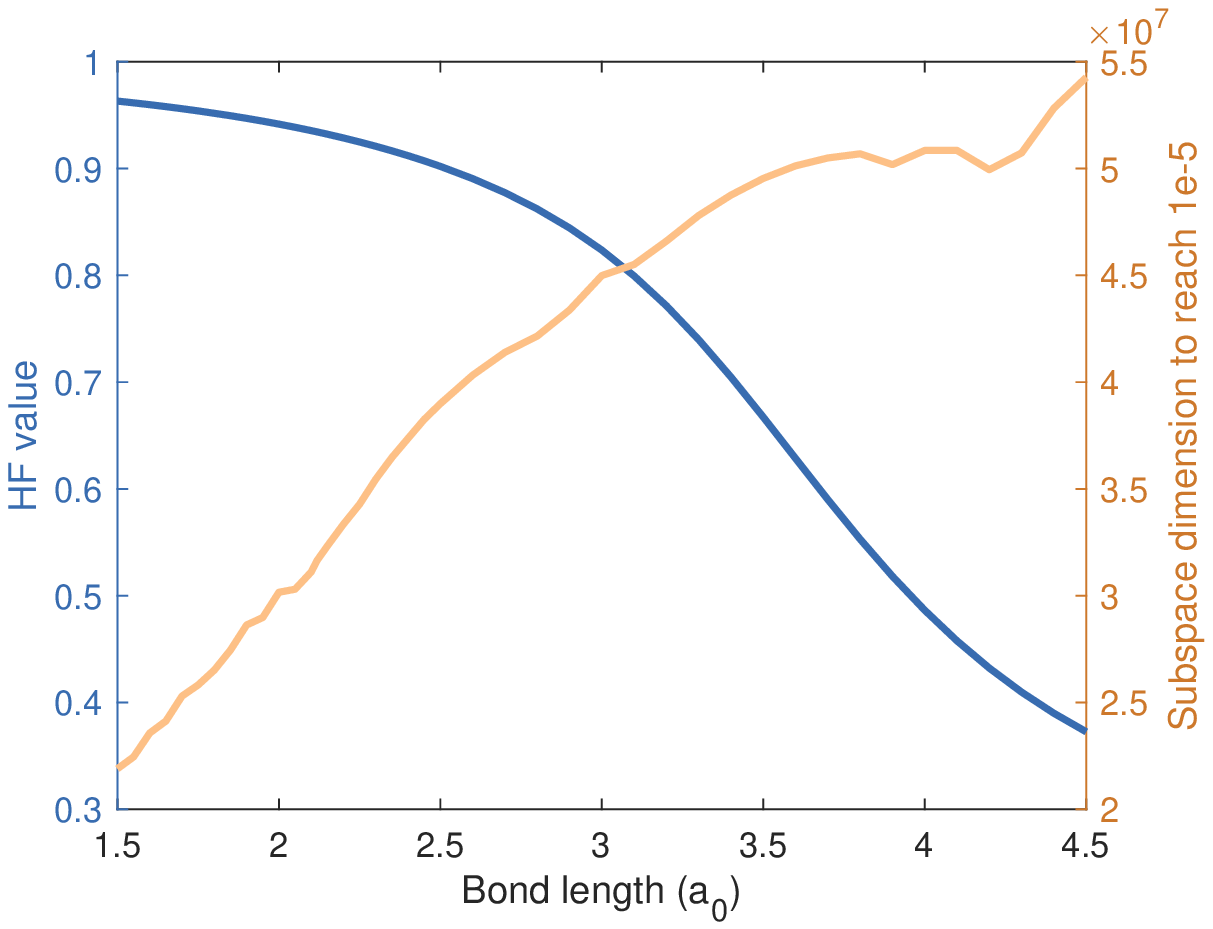}
    }
    \caption{Binding curve and the corresponding Hartree-Fock values
    and number of nonzeros in $\vec{c}$ of \ch{N2} under cc-pVDZ basis.}
    \label{fig:n2_binding_curve}
\end{figure}

In this problem, there are $14$ electrons and $28$ orbitals,
and the dimension of the FCI space is about
$N_{\text{FCI}} \approx 1.75\times 10^{11}$. In all configurations,
$\varepsilon = 10^{-6}$ is used in \sname{} for truncation.  Here we
use the same computing environment as in
Section~\ref{sec:num-res-molecule} but with OpenMP enabled with 5
threads. Each configuration on the bind curve results take roughly one
day to achieve the $10^{-3}$ mHa
accuracy. Figure~\ref{fig:n2_binding_curve} shows the binding curve
and Table~\ref{tab:n2_binding_curve1} and \ref{tab:n2_binding_curve2} in
Appendix~\ref{app:n2_binding_curve} list all converged variational
energies for every configuration in the figure.  In
Table~\ref{tab:n2_comparison}, we compare selected results obtained
from \sname{} with that from other algorithms reported in
Ref.~\citenum{Chan2004b}.

\begin{table*}
    \centering
    \begin{tabular}{ccccccc}
    \toprule
    Bond Length & 2.118$a_0$ & 2.4$a_0$ & 2.7$a_0$ & 3.0$a_0$
    & 3.6$a_0$ & 4.2$a_0$ \\
    \midrule
    \sname{} & -109.282173\wds{} & -109.241908\wds{} & -109.163600\wds{}
    & -109.089405\wds{} & -108.998083\wds{} & -108.970132\wds{} \\
    DMRG     & -109.2821\wds{57} & -109.2418\wds{86} & -109.1635\wds{72}
    & -109.0893\wds{80} & -108.9980\wds{52} & -108.9700\wds{90} \\
    CCSD     & -109.2\wds{67626} & -109.2\wds{19794} & -109.1\wds{31491}
    & -109.0\wds{52884} & -108.9\wds{75885} & -108.9\wds{60244} \\
    CCSDTQ   & -109.281\wds{943} & -109.241\wds{321} & -109.16\wds{2264}
    & -109.08\wds{6502} & -108.99\wds{3736} & -108.96\wds{8124} \\
    MRCISD   & -109.27\wds{5356} & -109.2\wds{34925} & -109.15\wds{6473}
    & -109.08\wds{2149} & -108.99\wds{0759} & -108.9\wds{63070} \\
    MRCCSD   & -109.28\wds{0646} & -109.24\wds{0362} & -109.16\wds{1969}
    & -109.08\wds{7613} & -108.99\wds{5885} & -108.96\wds{7865} \\
    \bottomrule
    \end{tabular}
    \caption{Nitrogen molecule ground state energy using \sname{}, DMRG
    ($\max M = 4000$) and couple cluster theories. \emph{Slant digits}
    indicate inaccurate digits. All results except \sname{} are from
     Ref.~\citenum{Chan2004b}.} \label{tab:n2_comparison}
\end{table*}

Several remarks are in order regarding the benchmark results.
First, Figure~\ref{fig:n2_binding_curve} demonstrates a smooth
standard shape binding curve. Different from the carbon binding
curve~\cite{Holmes2016}, no jump is observed in our benchmark results,
where $D_{2h}$ symmetry is used for all configurations. Second,
Table~\ref{tab:n2_comparison} shows that \sname{} gives the lowest
energy. \sname{} energy is accurate beyond the level of $10^{-3}$
mHa whereas DMRG is accurate up to $10^{-2}$ mHa. The DMRG
results in Table~\ref{tab:n2_comparison} are taken from previous
work~\cite{Chan2004b}, which agree with the results obtained in
Section~\ref{sec:c2n2}. Other algorithms such as CCSDTQ are much less
accurate. Third, the more the nitrogen dimer molecule is stretched
from the equilibrium, the more determinants and iterations are needed
for \sname{} to converge to $10^{-3}$ mHa accuracy. This is because
Hartree-Fock theory only works well near equilibrium configuration.
However, the number of determinants and iterations do not increase
significantly, which shows the efficiency of \sname{} again.
Other algorithms become less accurate for larger stretching distance.

\subsection{All electron chromium dimer calculation}
\label{sec:cr2}

Chromium dimer is hard to compute due to its strong correlation. We
calculate the all-electron molecule using \sname{} under the
Ahlrichs VDZ basis with radius $r = 1.5$\r{A}. There are $48$
electrons and $42$ orbitals, and the dimension of the FCI space
is about $2\times 10^{22}$. Many methods have been applied
to this problem including DMRG~\cite{Olivares-Amaya2015} and
HCI~\cite{Holmes2016}. Table~\ref{tab:cr2} summarizes all results,
including our \sname{} results and others from
literature~\cite{Olivares-Amaya2015, Holmes2016}.

\begin{table}
    \centering
    \begin{tabular}{cl}
    \toprule
    Algorithm & Energy (Ha)\\
    \midrule
        HCI (variational)    & $-2086.\wds{384}$    \\
        CCSD(T)              & $-2086.4\wds{22229}$ \\
        CCSDTQ               & $-2086.4\wds{30244}$ \\
        DMRG ($\max M=8000$)  & $-2086.44\wds{3334}$ \\
        \sname{}             & $-2086.443\wds{565}$  \\
    \midrule
        HCI (perturbed)      & $-2086.444\wds{04}$  \\
        DMRG (extrapolated)  & $-2086.444784$ \\
    \bottomrule
    \end{tabular}
    \caption{Energy of \ch{Cr2}.}
    \label{tab:cr2}
\end{table}

In this paper, we only consider variational ground state energy
without any perturbation or extrapolation. Regarding the variational
ground energy, \sname{} achieves lowest energy among all algorithms
in one month running time on a machine with Intel Xeon CPU E5-1650
v3 @ 3.50GHz and 128GB memory. Both DMRG ($\max M=8000$) and \sname{}
achieve the chemical accuracy if the energy of DMRG (extrapolated) is
regarded as the ground truth. But HCI(variational) and coupled cluster
theory cannot achieve chemical accuracy. HCI converges to $-2086.384$ Ha
in about eight minutes~\cite{Holmes2016}, whereas \sname{} reaches the
same accuracy in about twenty minutes, although different computing
environments are used.  As the dimension of the Hamiltonian becomes
larger and the system becomes more correlated, HCI can only afford
storing the submatrix in the main memory with very limited number of
determinants and such a limited number cannot achieve higher accuracy
in the variational phase. With perturbation phase enabled, HCI can
achieve accuracy less than $1$ mHa.  Similar perturbation phase can
be adapted to \sname{} to further boost the accuracy or extend the
applicability of our algorithm to larger systems.

\section{Conclusion and discussion}
\label{sec:conc}

The proposed \fname{} (\sname{}) is an easy-to-use, accurate, and
efficient algorithm for full configuration interaction eigenvalue
problems of quantum many-body systems, especially for strongly correlated
systems. The only tuning parameter in \sname{}, $\varepsilon$, controls
the trade-off between memory cost and accuracy. Given the fixed amount
of memory, the ``close-to-optimal'' $\varepsilon$ can be determined
within a few minutes without waiting for convergent results. Hence,
we believe that \sname{} is one of the most easy-to-use algorithms
among competitors. Besides the user friendly property, \sname{}
performs competitively with many other methods, including heat-bath
configuration interaction (HCI), density matrix renormalization group
(DMRG), and full configuration interaction quantum Monte Carlo with
initiator and semi-stochastic adaptation~(\fciqmc{}).  The \sname{} can
give the state-of-art results for many strongly correlated FCI problems.

There are a few immediate future work of \sname{}. Apply \sname{} to
current examples with larger basis sets and other more challenging
systems; add perturbation stage to further improve the accuracy;
and, parallelize \sname{} in a distributed-memory setting.
The earlier two can be accomplished directly, while the massive
distributed-memory parallelization of \sname{} requires modification
of the greedy determinant-select strategy. Hence FCIQMC type methods
currently have some advantage in that regard. As discussed in our
previous work~\cite{Li2018b}, with a stochastic variant of the
current determinant-select strategy, the asynchronized feature of
coordinate descent methods can be enabled.  \footnote{While it is
called asynchronized parallelization in coordinate descent methods,
communication is still needed after every several iterations. Hence
the embarrassing parallelization of Monte Carlo methods, as in FCIQMC
type methods, is of better scalability.} Massive distributed-memory
parallelized \sname{} is expected to achieve good performance. Besides
these, we are also exploring (semi-)stochastic \sname{} to improve
the parallelizability of the algorithm, and further accelerating
the initial iterations. Replacing the current hash function with
a more efficient one to fully utilize the memory hierarchy is also
under investigation. It is also interesting to design an auto-tuning
procedure for ``close-to-optimal'' $\varepsilon$ to remove the only
tuning parameter in the algorithm.

Beyond ground state computation, \sname{} is also suitable for
excited state computation.  The extension of the optimization problem
\eqref{eq:opt} to low-lying $k$ excited states can be achieved without
orthogonality constraint, i.e.,
\begin{equation} \label{eq:opt-k}
    \min_{\vec{c} \in \bbR^{N_{\text{FCI}} \times k}} f(\vec{c}) =
    \norm{H + \vec{c}\vec{c}^\top}_F^2.
\end{equation}
This is favorable as it removes the expensive orthogonalization step
for FCI wavefunctions during iterations.  Hence extending \sname{}
to solve low-lying excited states is another promising future
direction to be explored.
  
 
\begin{acknowledgement}
    The authors thank Qiming Sun for helpful discussions regarding
    PySCF and FCI calculations, Ali Alavi for insightful suggestions
    and for providing all configurations of NECI, and George Booth for
    helpful discussion on FCIQMC. The work is supported in part by the US
    National Science Foundation under awards DMS-1454939 and OAC-1450280
    and by the US Department of Energy via grant DE-SC0019449.
\end{acknowledgement}

% \begin{suppinfo}

% A listing of the contents of each file supplied as Supporting Information
% should be included. For instructions on what should be included in the
% Supporting Information as well as how to prepare this material for
% publications, refer to the journal's Instructions for Authors.

% The following files are available free of charge.

% %\begin{itemize}
% %    \item Filename: brief description
% %    \item Filename: brief description
% %\end{itemize}

% \end{suppinfo}

\nocite{*}
\bibliography{reference}

%%% Add Appendix
\newpage
\appendix

\section{\ch{N2} Binding Curve}\label{app:n2_binding_curve}

Figure~\ref{fig:n2_binding_curve} plots the binding
curve of nitrogen dimer in cc-pVDZ basis with
data given in Table~\ref{tab:n2_binding_curve1}
and Table~\ref{tab:n2_binding_curve2}.  The bond
length of nitrogen dimer in equilibrium geometry
is $2.118 a_0$. Table~\ref{tab:n2_binding_curve1} and
Table~\ref{tab:n2_binding_curve2} list variational energies of
nitrogen dimer produced by \sname{} with bond lengths smaller and
larger than $2.118 a_0$ respectively. In both tables, \sname{} used
$\varepsilon = 10^{-6}$ as the truncation threshold.

The bond lengths are selected through the following two steps. \sname{}
first calculates energies for a vector of bond lengths linearly spaced
between and including $1.50 a_0$ and $4.50 a_0$ with gap $0.10 a_0$.
Then, according to the initial rough binding curve, another vector of
bond lengths is added to smooth out the curve. These added bond lengths
are in the sharp changing range around the equilibrium setting.

\begin{table}[H]
\centering
\begin{tabular}{cc}
\toprule
Bond length ($a_0$) & Energy (Ha) \\
\midrule
$1.50$ & $-108.6300476$  \\
$1.55$ & $-108.7719968$  \\
$1.60$ & $-108.8888460$  \\
$1.65$ & $-108.9843136$  \\
$1.70$ & $-109.0615754$  \\
$1.75$ & $-109.1233484$  \\
$1.80$ & $-109.1719641$  \\
$1.85$ & $-109.2094264$  \\
$1.90$ & $-109.2374578$  \\
$1.95$ & $-109.2575411$  \\
$2.00$ & $-109.2709530$  \\
$2.05$ & $-109.2787896$  \\
$2.10$ & $-109.2819938$  \\
$2.118$ & $-109.2821727$  \\
\bottomrule
\end{tabular}
\caption{Energy of nitrogen dimer with bond length smaller than that
of equilibrium geometry. The energy refers to variational ground
state energy calculated by \sname{} with $\varepsilon = 10^{-6}$.}
\label{tab:n2_binding_curve1}
\end{table}

\begin{table}[H]
\centering
\begin{tabular}{cc}
\toprule
Bond length ($a_0$) & Energy (Ha) \\
\midrule
$2.118$ & $-109.2821727$  \\
$2.15$ & $-109.2813737$  \\
$2.20$ & $-109.2776211$  \\
$2.25$ & $-109.2713283$  \\
$2.30$ & $-109.2630013$  \\
$2.35$ & $-109.2530718$  \\
$2.40$ & $-109.2419079$  \\
$2.45$ & $-109.2298228$  \\
$2.50$ & $-109.2170830$  \\
$2.60$ & $-109.1905077$  \\
$2.70$ & $-109.1635998$  \\
$2.80$ & $-109.1373583$  \\
$2.90$ & $-109.1124729$  \\
$3.00$ & $-109.0894053$  \\
$3.10$ & $-109.0684502$  \\
$3.20$ & $-109.0497787$  \\
$3.30$ & $-109.0334619$  \\
$3.40$ & $-109.0194835$  \\
$3.50$ & $-109.0077466$  \\
$3.60$ & $-108.9980829$  \\
$3.70$ & $-108.9902691$  \\
$3.80$ & $-108.9840499$  \\
$3.90$ & $-108.9791625$  \\
$4.00$ & $-108.9753572$  \\
$4.10$ & $-108.9724102$  \\
$4.20$ & $-108.9701316$  \\
$4.30$ & $-108.9683664$  \\
$4.40$ & $-108.9669909$  \\
$4.50$ & $-108.9659102$  \\
\bottomrule
\end{tabular}
\caption{Energy of nitrogen dimer with bond length larger than that
of equilibrium geometry. The energy refers to variational ground
state energy calculated by \sname{} with $\varepsilon = 10^{-6}$.}
\label{tab:n2_binding_curve2}
\end{table}

\end{document}